\documentclass[runningheads]{llncs}
\usepackage[T1]{fontenc}
\usepackage{graphicx}
\usepackage{tabularx}
\usepackage{array,booktabs,ragged2e}
\newcolumntype{R}[1]{>{\RaggedLeft\arraybackslash}p{#1}}
\begin{document}
\title{Automated CT Lung Cancer Screening Workflow using 3D Camera}

\titlerunning{Automated CT Lung Cancer Screening}
%
\author{Brian Teixeira\inst{1} \and
Vivek Singh\inst{1} \and
Birgi Tamersoy\inst{2} \and
Andreas Prokein\inst{3} \and
Ankur Kapoor\inst{1}}


%
\authorrunning{B. Teixeira et al.}


%
\institute{Digital Technology \& Innovation, Siemens Healthineers, Princeton, NJ, USA, \and
Digital Technology \& Innovation, Siemens Healthineers, Erlangen, Germany \and
Computed Tomography, Siemens Healthineers, Forchheim, Germany}

\maketitle              
\begin{abstract}
Despite recent developments in CT planning that enabled automation in patient positioning, time-consuming scout scans are
still needed to compute dose profile and ensure the patient is properly positioned.
In this paper, we present a novel method which eliminates the need for scout scans in CT lung cancer screening by estimating patient scan range, isocenter, and Water Equivalent Diameter (WED) from 3D camera images.
We achieve this task by training
an implicit generative model on over 60,000 CT scans and introduce a novel approach for updating the prediction
using real-time scan data. We demonstrate the effectiveness of our method on a testing set of 110 pairs
of depth data and CT scan, resulting in an average error of $5 mm$ in estimating the isocenter, $13 mm$ in determining
the scan range, $10 mm$ and $16 mm$ in estimating the AP and lateral WED respectively. The relative WED error of our method is $4\%$, which is well within the International Electrotechnical Commission (IEC) acceptance criteria of $10\%$.

\end{abstract}
\section{Introduction}

Lung cancer is the leading cause of cancer death in the United States, and early detection is key to
improving survival rates. CT lung cancer screening is a low-dose CT (LDCT) scan of the chest that can detect lung cancer
at an early stage, when it is most treatable. However, the current workflow for performing CT lung scans
still requires an experienced technician to manually perform pre-scanning steps, which greatly decreases
the throughput of this high volume procedure. While recent advances in human body modeling  \cite{darwin,united,brian_cvpr18,osso,miccai18} have allowed for
automation of patient positioning, scout scans are still required as they are used by automatic exposure control system in the CT scanners to compute the dose to be delivered in order to maintain constant image quality \cite{aec}.

Since LDCT scans are obtained in a single breath-hold and do not require any contrast medium to be injected, the scout scan consumes a significant portion of the scanning workflow time. It is further increased by the fact that tube rotation has to be adjusted between the scout and actual CT scan.
Furthermore, any patient movement during the time between the two scans may cause misalignment and
incorrect dose profile, which could ultimately result in a repeat of the entire process. Finally, while
minimal, the radiation dose administered to the patient is further increased by a scout scan.

We introduce a novel method for estimating patient scanning
parameters from non-ionizing 3D camera images to eliminate the need for scout scans during pre-scanning. For LDCT lung cancer screening, our framework  automatically estimates the patient's lung position (which serves as a reference point
to start the scan), the patient's isocenter (which is used to determine the table height for scanning), and an estimate of patient's Water
Equivalent Diameter (WED) profiles along the craniocaudal direction which is a well established method for defining Size Specific Dose Estimate (SSDE)
in CT imaging \cite{aapm_wed,wed_chest,wed_radio,ssde_myo}. Additionally, we introduce a novel
approach for updating the estimated WED in real-time, which allows for refinement of the scan parameters during acquisition, thus increasing accuracy. We present a method for automatically aborting the scan if the predicted WED deviates from real-time acquired data beyond the clinical limit. We trained our models on a large collection of CT scans acquired from
over $60,000$ patients from over 15 sites across North America, Europe and Asia. The contributions of this work can
be summarized as follows:

\begin{itemize}
  \item A novel workflow for automated CT Lung Cancer Screening without the need for scout scan
  \item A clinically relevant method meeting IEC 62985:2019 requirements on WED estimation.
  \item A generative model of patient WED trained on over $60,000$ patients.
  \item A novel method for real-time refinement of WED, which can be used for dose modulation
\end{itemize}
\section{Method}
\begin{figure*}[ht]
  \begin{center}
  \centering
  \includegraphics[width=1\textwidth]{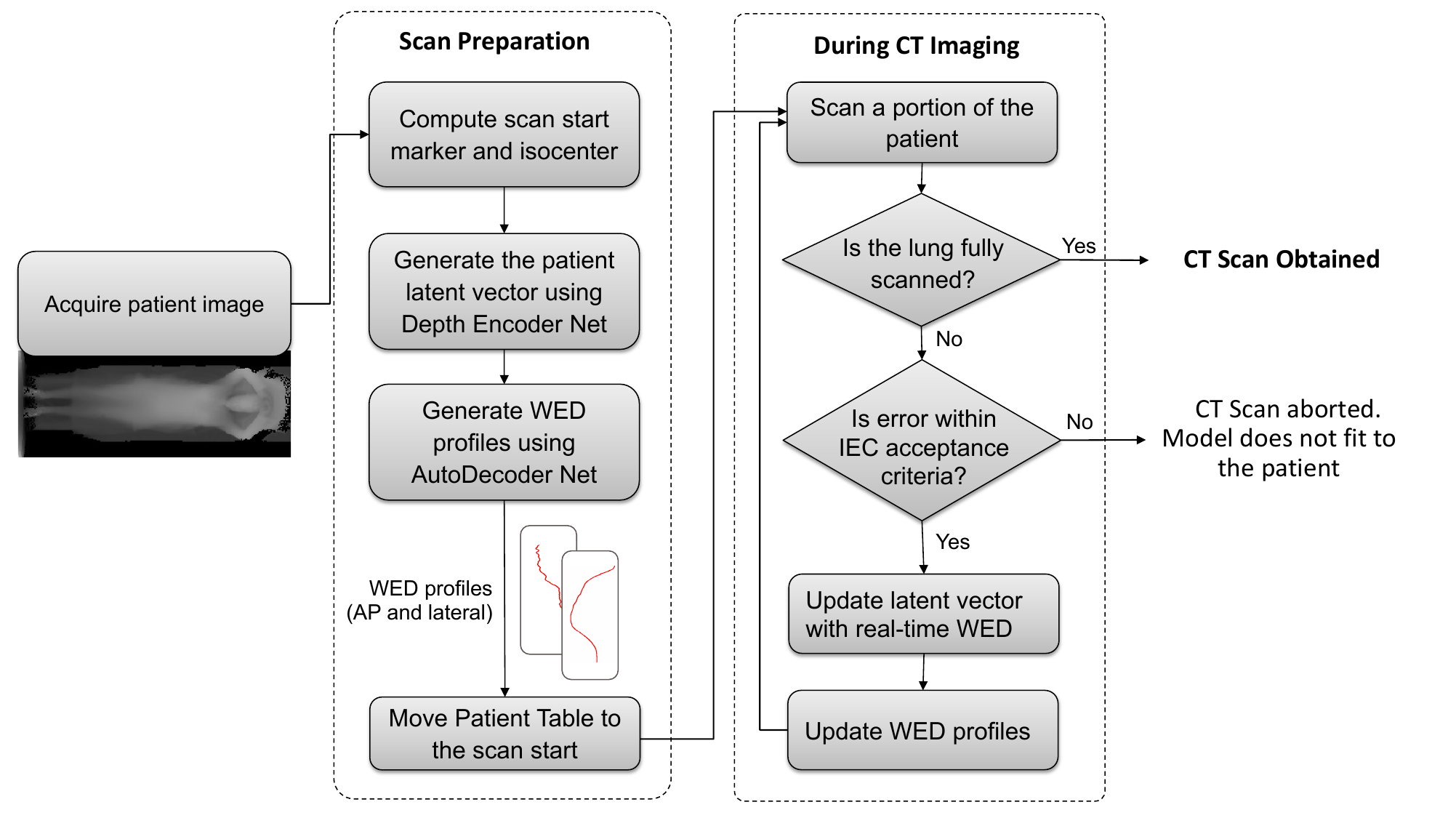}
  \caption{Overview of the proposed workflow.
  }
    \label{fig:flowchart}
  \end{center}
\end{figure*}

Water Equivalent Diameter (WED) is a robust patient-size descriptor \cite{wed_loc} used for CT dose planning.
It represents the diameter of a cylinder of water having the same averaged absorbed dose as the material contained
in an axial plane at a given craniocaudal position $z$ \cite{IEC62985}. The WED of a patient is thus a function
taking as input a craniocaudal coordinate and outputting the WED of the patient at that given position. As WED is
defined in an axial plane, the diameter needs to be known on both the Anterior-Posterior (AP) and lateral (Left-
Right) axes noted respectively $WED_{AP}(z)$ and $WED_{L}(z)$. As our focus here is on lung cancer screening, we define `WED profile' to be the 1D curve obtained by
uniformly sampling the WED function along the craniocaudal axis
within the lung region. Our method jointly predicts the AP and lateral WED profiles.

\vspace{0.5mm}

While WED can be derived from CT images, paired CT scans and camera images are rarely available, making
direct regression through supervised learning challenging. We propose a semi-supervised approach to estimate WED from depth images. First, we train a WED generative model
on a large collection of CT scans. We then train an encoder network to map
the patient depth image to the WED manifold. Finally, we propose a novel method to refine the
prediction using real-time scan data.

\subsection{WED Latent Space Training}

\vspace{0.5mm}

We use an AutoDecoder \cite{deepsdf} to learn the WED latent space. Our model is a fully connected network with 8 layers of 128 neurons each.
We used layer normalization and ReLU activation after each layer except the last one.
Our network takes as input a latent vector together with a craniocaudal coordinate $z$
and outputs $WED_{AP}(z)$ and $WED_{L}(z)$, the values of the AP and lateral WED at the given coordinate. In this approach, our latent vector
represents the encoding of a patient in the latent space. This way, a single AutoDecoder can learn patient-specific continuous WED functions. Since our network only takes the craniocaudal
coordinate and the latent vector as input, it can be be trained on partial scans of different sizes. The
training consists of a joint optimization of the AutoDecoder and the latent vector: the AutoDecoder is
learning a realistic representation of the WED function while the latent vector is updated to
fit the data.

\vspace{0.5mm}

During training, we initialize our latent space to a unit Gaussian distribution as we want it
to be compact and continuous. We then randomly sample points along the craniocaudal axis and minimize the L1 loss
between the prediction and the ground truth WED. We also apply L2-regularization on the latent vector as part of the optimization process.

\subsection{Depth Encoder Training}

\vspace{0.5mm}

After training our generative model on a large collection of unpaired CT scans, we train our encoder
network on a smaller collection of paired depth images and CT scans. We represent our encoder as a
DenseNet \cite{densenet} taking as input the depth image and outputting a
latent vector in the previously learned latent space. Our model has 3 dense blocks of 3 convolutional layers.
Each convolutional layer (except the last one) is followed by a spectral normalization layer and a ReLU activation.
The predicted latent vector is then used as input to the frozen AutoDecoder to generate the predicted
WED profiles. We here again apply L2-regularization on the latent vector during training.

\vspace{0.5mm}

\subsection{Real-time WED Refinement}

While the depth image provides critical information on the patient anatomy, it may not always be
sufficient to accurately predict the WED profiles. For example, some patients may have implants
or other medical devices that cannot be guessed solely from the depth image. Additionally, since the
encoder is trained on a smaller data collection, it may not be able to perfectly project the
depth image to the WED manifold. To meet the strict safety criteria defined by the IEC, we propose to dynamically
update the predicted WED profiles at inference time using real-time scan data. First, we use our encoder
network to initialize the latent vector to a point in the manifold that is close to the
current patient. Then, we use our AutoDecoder to generate initial WED profiles. As the table moves
and the patient gets scanned, CT data is being acquired and ground truth WED can be computed for
portion of the body that has been scanned, along with the corresponding craniocaudal coordinate. We can then
use this data to optimize the latent vector by freezing the AutoDecoder and minimizing the L1 loss between
the predicted and ground truth WED profiles through gradient descent. We can then feed the updated latent vector
to our AutoDecoder to estimate the WED for the remaining portions of the body that have not yet been scanned and repeat the process.

\vspace{0.5cm}

In addition to improving the accuracy of the WED profiles prediction, this approach can also help detect deviation
from real data. After the latent vector has been optimized to fit the previously scanned data, a large deviation between
the optimized prediction and the ground truth profiles may indicate that our approach is not able to
find a point in the manifold that is close to the data. In this case, we may abort the scan, which further reduces safety risks. Overall flowchart of the proposed approach is shown in Figure
\ref{fig:flowchart}.
\section{Results}
\begin{figure*}[h]
  \begin{center}
  \centering
  \includegraphics[width=1\textwidth]{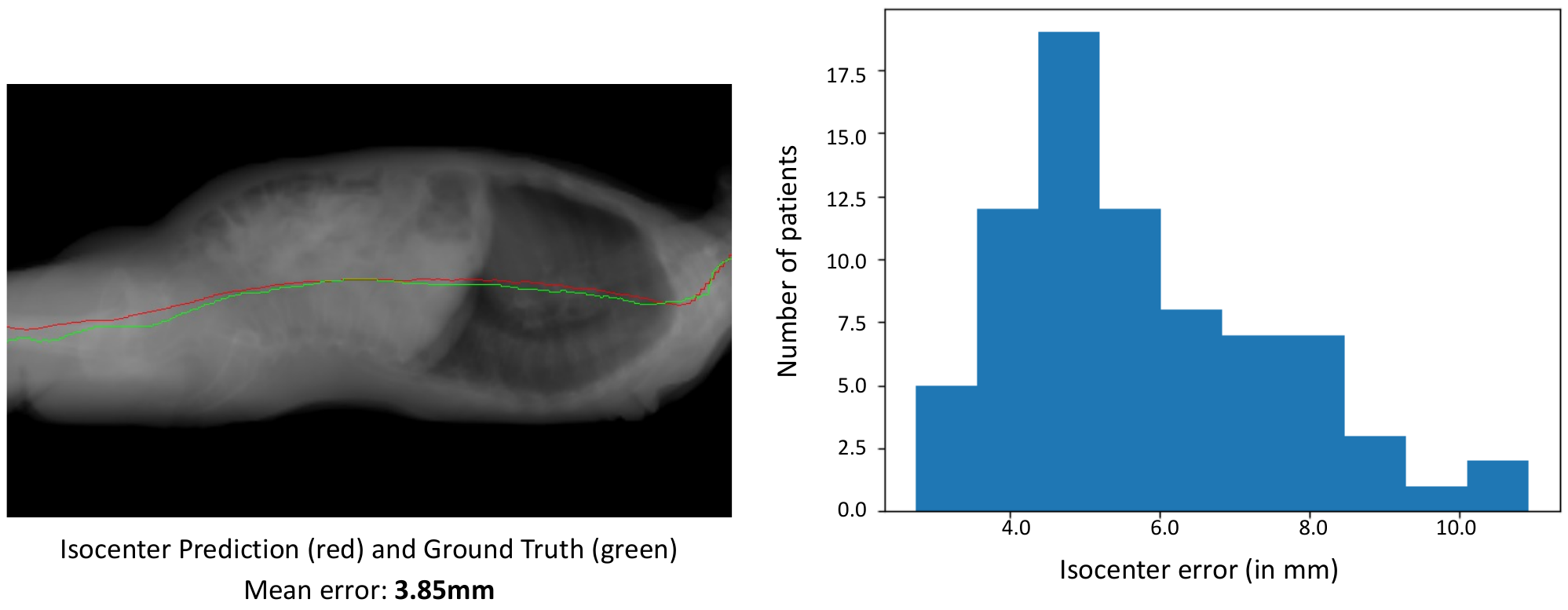}
  \caption{Isocenter results on our evaluation set. Left column presents a qualitative result from our evaluation set.
          The red line corresponds to our model prediction and the green line is the ground truth computed from the CT.
          The right column presents a histogram of the errors in mm.
  }
    \label{fig:isocenter_results}
  \end{center}
\end{figure*}
\subsection{Data}
Our CT scan dataset consists of $62,420$ patients from $16$ different sites across North America, Asia and Europe.
Our 3D Camera dataset consists of $2,742$ pairs of depth image and CT scan from $2,742$ patients from $6$ different sites across North
America and Europe acquired using a ceiling-mounted Kinect 2 camera.
Our evaluation set consists of $110$ pairs of depth image and CT scan from $110$ patients from a separate site in Europe.

\subsection{Patient preparation}
Patient positioning is the first step in lung cancer screening workflow. We first need to estimate the table position and the starting point of the scan.
We propose to estimate the table position by regressing the patient isocenter and the starting point of the scan by estimating the location of the patient's lung top.

\subsubsection{Starting position}
We define the starting position of the scan as the location of the patient's lung top. We trained a DenseUNet \cite{denseunet} taking the camera depth image as input
and outputting a Gaussian heatmap centered at the patient's lung top location. We used 4 dense blocks of 4 convolutional layers for the encoder
and 4 dense blocks of 4 convolutional layers for the decoder. Each convolutional layer (except the last one) is followed by a batch normalization layer and a ReLU activation.
We trained our model on $2,742$ patients using Adaloss \cite{adaloss} and the Adam \cite{adam} optimizer
with a learning rate of $0.001$ and a batch size of $32$ for 400 epochs. Our model achieves a mean error of $\textbf{12.74mm}$ and a $95^{th}$ percentile error of $\textbf{28.32mm}$.
To ensure the lung is fully visible in the CT image, we added a $2cm$ offset on our
prediction towards the outside of the lung. We then defined the accuracy as whether
the lung is fully visible in the CT image when using the offset prediction.
We report an accuracy of $\textbf{100\%}$ on our evaluation set of $110$ patients.

\subsubsection{Isocenter}

The patient isocenter is defined as the centerline of the patient's body. We trained a DenseNet \cite{densenet} taking the camera depth image as input and outputting
the patient isocenter. Our model is made of 4 dense blocks of 3 convolutional layers.
Each convolutional layer (except the last one) is followed by a batch normalization layer and a ReLU activation.
We trained our model on $2,742$ patients using Adadelta \cite{adadelta} with a batch size of $64$ for 300 epochs.
On our evaluation set, our model outperforms the technician's estimates with a mean error of $\textbf{5.42mm}$ and a $95^{th}$ percentile error of $\textbf{8.56mm}$ compared to $6.75$mm and $27.17$mm
respectively. Results can be seen in Figure \ref{fig:isocenter_results}.

\subsection{Water Equivalent Diameter}
\begin{figure*}[t]
  \begin{center}
  \centering
  \includegraphics[width=1\textwidth]{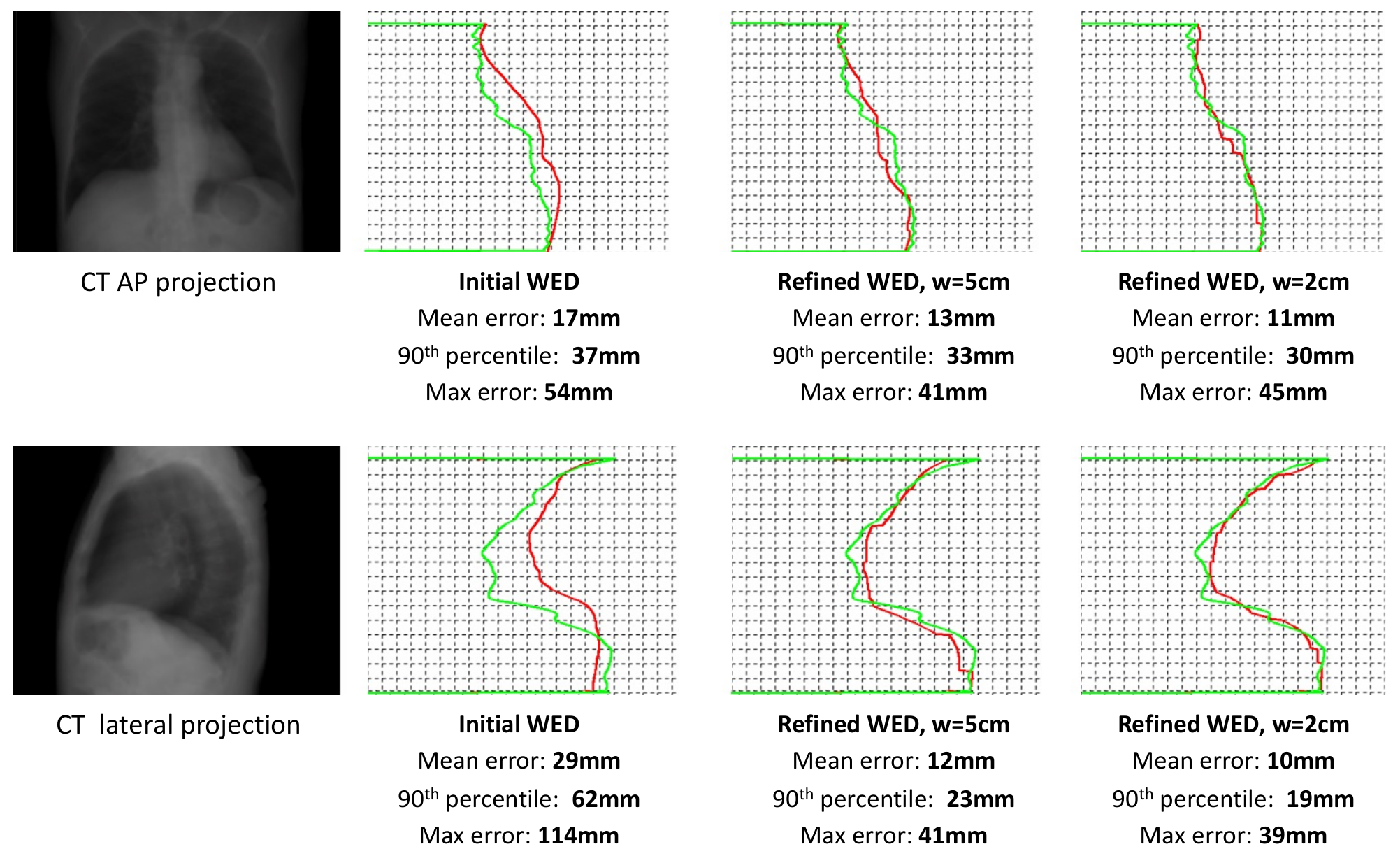}
  \caption{AP (top) and lateral (bottom) WED profile regression with and without real-time refinement.
              $w$ corresponds to the portion size of the body that gets scanned before updating the prediction (in cm).
           First column shows a lateral projection view of the CT. Second column shows the performance
           of our model without real-time refinement. Third and fourth columns show the
performance of our model with real-time refinement every 5cm and 2 cm respectively. Ground truth is depicted in green
           and our prediction is depicted in red. While the original prediction was off towards the center
           of the lung, the real-time refinement was able to correct the error.
  }
    \label{fig:both_attenuation_results}
  \end{center}
\end{figure*}

\begin{table}[t]
\caption{WED profile errors on our testing set (in $mm$).
         `w' corresponds to the portion size of the body that gets scanned before updating the prediction (in cm).
         Top of the table corresponds to lateral WED profile, bottom corresponds to AP WED profile.
         Updating the prediction every 20mm produces the best results.}\label{table_att}
\begin{tabularx}{\textwidth}{l|R{2cm}|R{3cm}|R{2.75cm}}
\hline
\textbf{Method (lateral)} & \textbf{Mean error} & \textbf{90th perc error} & \textbf{Max error}\\
\hline
\textit{Direct Regression} & $45.07$ & $76.70$ & $101.50$\\
\hline
\textit{Proposed (initial)} & $27.06$ & $52.88$ & $79.27$\\
\hline
\textit{Proposed (refined, $w=5$)} & $19.18$ & $42.44$ & $73.69$\\
\hline
\textit{Proposed (refined, $w=2$)} & $\textbf{15.93}$ & $\textbf{35.93}$ & $\textbf{61.68}$\\
\hline
\textbf{Method (AP)} &  &  & \\
\hline
\textit{Direct Regression} & $45.71$ & $71.85$ & $82.84$\\
\hline
\textit{Proposed (initial)} & $16.52$ & $31.00$ & $40.89$\\
\hline
\textit{Proposed (refined, $w=5$)} & $12.19$ & $25.73$ & $37.36$\\
\hline
\textit{Proposed (refined, $w=2$)} & $\textbf{10.40}$ & $\textbf{22.44}$ & $\textbf{33.85}$\\
\hline
\end{tabularx}
\end{table}

We trained our AutoDecoder model on our unpaired CT scan dataset of $62,420$ patients with a latent vector of size 32. The encoder was trained on our paired CT scan and depth image dataset of $2,742$ patients.
We first compared our method against a simple direct regression model. We trained a DenseUNet \cite{denseunet} taking the camera depth image as input and outputting the Water Equivalent Diameter profile.
We trained this baseline model on $2,742$ patients using the Adadelta \cite{adam} optimizer with a learning rate of $0.001$ and a batch size of $32$. We then measured the performance of our model
before and after different degrees of real-time refinement, using the same optimizer and learning rate.  We report the comparative results in Table \ref{table_att}.

We observed that our method largely outperforms
the direct regression baseline with a mean lateral error $\textbf{40\%}$ lower and a $90^{th}$ percentile lateral error over $\textbf{30\%}$ lower. Bringing in real-time refinement greatly improves the results with a mean lateral error
over $\textbf{40\%}$ and a $90^{th}$ percentile lateral error over $\textbf{20\%}$ lower than before refinement. AP profiles
show similar results with a mean AP error improvement of nearly $\textbf{40\%}$ and a $90^{th}$ percentile AP error improvement close to $\textbf{30\%}$. When using our proposed method with a $20mm$ window refinement, our proposed approach outperforms the direct regression baseline
by over $\textbf{60\%}$ for lateral profile and nearly $\textbf{80\%}$ for AP.

\begin{figure*}[t]
  \begin{center}
  \centering
  \includegraphics[width=\textwidth]{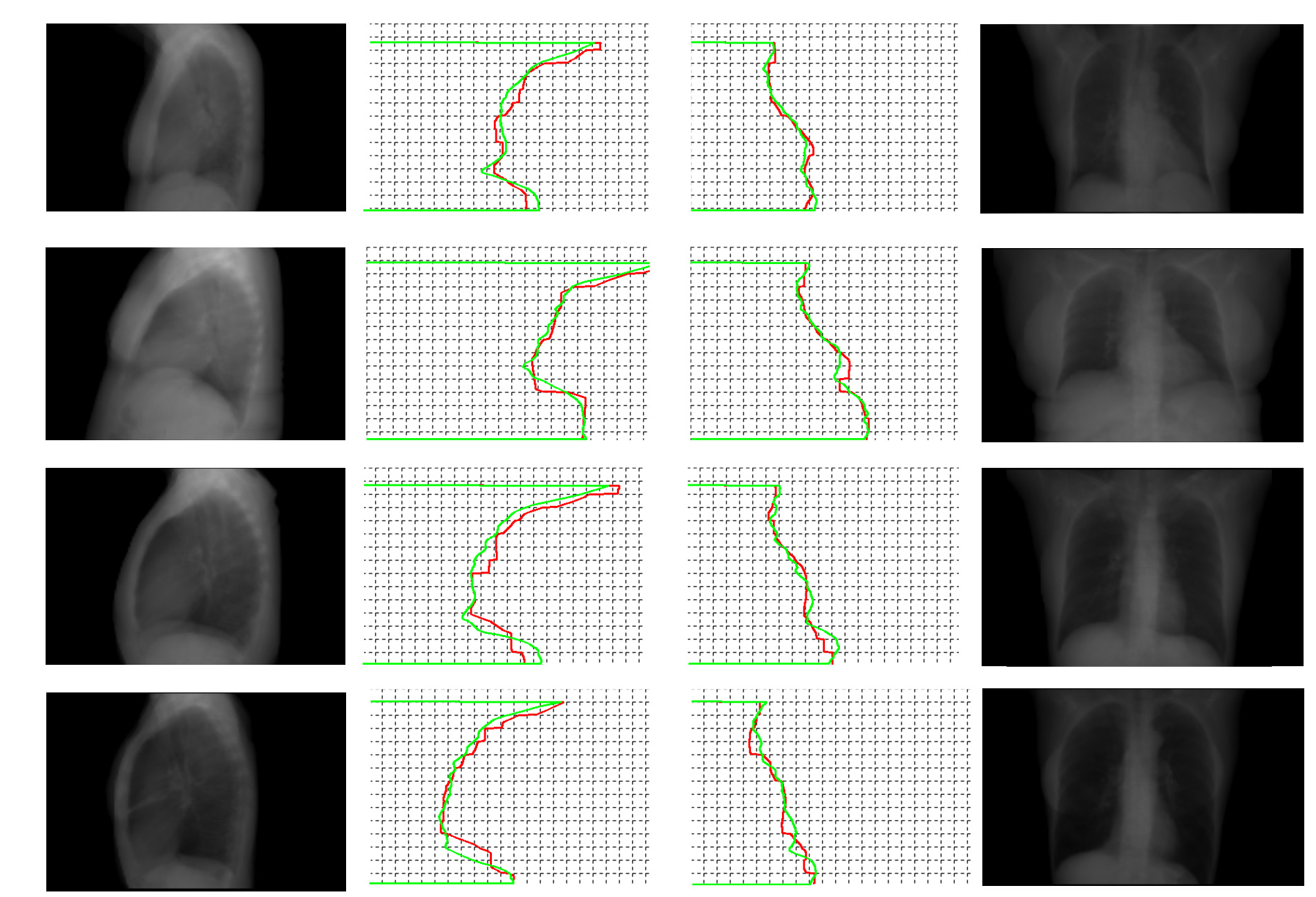}
  \caption{Qualitative analysis of the proposed method with 2cm refinement on patient with different morphology. From left to right: Lateral CT projection, Lateral WED profile, AP WED profile, AP CT projection.
  }
    \label{fig:many_attenuations}
  \end{center}
\end{figure*}

Figures \ref{fig:both_attenuation_results} highlights the benefits of using real-time refinement. Overall,
our approach shows best results with an update frequency of $20$mm, with a mean lateral error of $\textbf{15.93mm}$ and a mean AP error
of $\textbf{10.40mm}$. Figure \ref{fig:many_attenuations} presents a qualitative evaluation
on patients with different body morphology.

Finally, we evaluated the clinical relevancy of our approach by computing the relative error as described in the
International Electrotechnical Commission (IEC) standard IEC 62985:2019 on \textit{Methods for calculating size specific dose estimates (SSDE) for computed tomography} \cite{IEC62985}. The $\Delta_{REL}$ metric is defined as:

\begin{equation}
\Delta_{REL}(z) = \left|\frac{\hat{WED}(z) - WED(z)}{WED(z)}\right|
\end{equation}

Where:
\begin{itemize}
  \item $\hat{WED}(z)$ is the predicted water equivalent diameter
  \item $WED(z)$ is the ground truth water equivalent diameter
  \item $z$ is the position along the craniocaudal axis of the patient.
\end{itemize}

IEC standard states the median value of the set of $\Delta_{REL}(z)$ along the craniocaudal axis (noted $\Delta_{REL}$) should be below $\textbf{0.1}$.
Our method achieved a mean lateral $\Delta_{REL}$ error of $\textbf{0.0426}$ and a mean AP $\Delta_{REL}$ error of $\textbf{0.0428}$, falling well within the acceptance criteria.

\section{Conclusion}
We presented a novel 3D camera based approach for automating CT lung cancer screening workflow
without the need for a scout scan.
Our approach effectively estimates start of scan, isocenter and
Water Equivalent Diameter from depth images and meets the IEC acceptance criteria of relative WED error.
While this approach can be used for other thorax scan protocols, it may not be applicable to trauma
(e.g. with large lung resections) and inpatient settings, as the deviation in predicted and actual WED
would likely be much higher. In future, we plan to establish the feasibility as well as the utility of
this approach for other scan protocols and body regions. \footnote{Disclaimer: The concepts and information presented in this paper
are based on research results that are not commercially available. Future commercial availability cannot be guaranteed.}

%
%
\bibliographystyle{splncs04}
\bibliography{paper3582}
\end{document}